\documentclass[aps]{revtex4}

\def\be{\begin{eqnarray}}
\def\beq{\begin{equation}}
\def\eeq{\end{equation}}
\def\ee{\end{eqnarray}}

\begin{document}
%\draft
\title  {\bf Pairing interactions and vanishing pairing correlations in hot nuclei}

\author{E. Khan$^{a)}$, Nguyen Van Giai$^{a)}$, N. Sandulescu$^{a),b)}$}

\vspace {03mm}

\address{
{\it a) Institut de Physique Nucl\'eaire, Universit\'e Paris-Sud, IN$_{2}$P$_{3}$-CNRS,
91406 Orsay Cedex, France}\\
{\it b) Institute for Physics and Nuclear Engineering, P.O. Box MG-6, 76900
Bucharest, Romania}\\
}

%\date{today}

\begin{abstract}
Finite temperature Hartree-Fock-Bogoliubov calculations are performed in Sn
isotopes using Skyrme and zero-range, density-dependent pairing interactions.
For both stable and very neutron-rich nuclei the critical temperature at
which pairing correlations vanish is independent of the volume/surface
nature of the pairing interaction. The value of the critical temperature
follows approximatively the empirical rule T$_c$ $\simeq$ 0.5 $\Delta_{T=0}$
for all the calculated isotopes, showing that the critical temperature could
be deduced from the pairing gap at zero temperature. On the other hand, the
pairing gap at temperatures just below T$_c$ is strongly sensitive to the
volume/surface nature of the pairing interaction.  
\end{abstract}

\vskip 0.5cm

%\twocolumn
\maketitle

\section{Introduction}

The competition between the temperature and pairing correlations in hot
nuclei has been studied for more than four decades. The first studies were
based on the BCS approximation \cite{sa63} but later on more involved
calculations based on the Bogoliubov approach have been performed
\cite{go81,eg93}. More recently the Bogoliubov approach has been employed
together with self-consistent Hartree-Fock mean fields in order to study the
pairing properties of hot nuclei. One of the first finite-temperature HFB
(FT-HFB) calculations was based on a finite-range force of Gogny type, which
is used for describing both the mean field and the pairing properties of hot
nuclei \cite{eg00}. FT-HFB calculations using zero-range forces have been
done for hot nuclei \cite{kh04} and for the inner crust matter of neutron
stars \cite{sa04}. In the latter case the mean field is obtained by using a
Skyrme force while the pairing correlations are calculated with a
density-dependent delta interaction. Also, shell-model approaches
\cite{ka04,la05} have been used in order to probe the impact of the
temperature on both pairing and deformation degrees of freedom.

The interplay between temperature and pairing correlations was also studied
intensively in nuclear and neutron matter \cite{de05,mu03}. Typically, the
pairing gaps are calculated in the BCS approximation and using
single-particle states determined by self-consistent Brueckner-Hartree-Fock
or Green's function methods (see \cite{ba99,mu05} and references therein).
In more fundamental approaches, which go beyond BCS approximation, is still
unclear how much the pairing correlations are affected by the medium
dependence of the nucleon-nucleon interaction (see \cite{ca06,go05} and
references therein).
 
One open issue in current HFB calculations is how much the form of the
pairing interaction affects the properties of nuclei, especially when one
approaches the drip lines. It is also not clear yet if one really needs to
introduce an explicit density dependence in the pairing interaction in order
to enforce a pairing field evenly distributed in the nucleus ("volume type
pairing") or strongly localized in the surface region ("surface type
pairing"). Since a realistic pairing force derived from first principles is
missing, one hopes to disentangle between various types of pairing forces by
analyzing their consequences on measurable quantities. However, up to now
these studies are not conclusive. For instance, in Ref. \cite{do01} a mixed
surface-volume pairing interaction is considered to better explain the
odd-even mass differences of some isotopic chains, whereas in Ref.
\cite{yu03} the surface or the volume type of the pairing interaction is
found to be not so relevant for the neutron separation energies. It is also
worth stressing that the pair density, which gives indications upon the
localization of pair correlations in finite nuclei, is not strongly
correlated to the surface or volume character of the pairing force but
rather to the localization of the single-particle states close to the
chemical potential \cite{sa05}.

Apart from the effects mentioned above, the type of the pairing force could
also affect the vanishing of pairing correlations in hot nuclei. Besides
constant G studies there have not been such systematic studies with
effective density-dependent pairing interactions. It is known that, in a
simple BCS approach with a constant pairing G, the vanishing of pairing
correlations is expected to occur at T$_c$$\simeq$0.5 $\Delta_{T=0}$
\cite{go81}. The aim of the present work is to analyze if the volume or
surface character of the pairing force could significantly influence
vanishing pairing correlations using density-dependent pairing interactions.
It should be noted that experimentally, the critical temperature could be
extracted from the change of the specific heat in the vanishing pairing correlations
region, using level densities measurements, as shown in Refs.
\cite{gu01,sc01}.

\section{Finite-Temperature Hartree-Fock-Bogoliubov with Skyrme interactions}

In this work we employ the FT-HFB approach with zero-range forces. Details
  can be found elsewhere \cite{kh04,sa04}, and we recall only the main
  equations. The FT-HFB equations, in coordinate representation, have the
  following form : \beq
\begin{array}{c}
\left( \begin{array}{cc}
h_T(r) - \lambda & \Delta_T (r) \\
\Delta_T (r) & -h_T(r) + \lambda 
\end{array} \right)
\left( \begin{array}{c} U_i (r) \\
 V_i (r) \end{array} \right) = E_i
\left( \begin{array}{c} U_i (r) \\
 V_i (r) \end{array} \right) ~,
\end{array}
\label{1} \eeq where $E_i$ is the quasiparticle energy, $U_i$, $V_i$ are 
the components of the radial FT-HFB wave function and $\lambda$ is the
chemical potential. The quantity $h_T(r)$ is the thermal averaged mean field
Hamiltonian and $\Delta_T (r)$ is the thermal averaged pairing field. The
latter is calculated with a density-dependent contact force of the following
form \cite{be91}: 

\beq V (\mathbf{r}-\mathbf{r^\prime}) = V_0 [1 -\eta 
(\frac{\rho(r)}{\rho_0})^{\alpha}] \delta(\mathbf{r}-\mathbf{r^\prime}) 
\equiv V_{eff}(\rho(r)) \delta(\mathbf{r}-\mathbf{r^\prime}) , 
\label{eqpaiint}
\eeq
where $\rho(r)$ is the baryonic density. With this force the thermal averaged
pairing field is local and is given by:

\begin{eqnarray}
\Delta_T(r) & = & \frac{1}{2}V_{eff}(\rho(r)){
\frac{1}{4\pi} \sum_{i} (2j_i+1) U_i^* (r) 
V_i (r)
(1 - 2f_i )} \nonumber \\
& \equiv & \frac{1}{2}V_{eff}(\rho(r)) \kappa_T (r) ,
\label{eqvpair}
\end{eqnarray}
where $\kappa_T(r)$ is the thermal averaged pairing tensor.
Due to the density dependence of the pairing force, the
thermal averaged mean field Hamiltonian $h_T(r)$ depends 
also on $\kappa_T$. In addition, the averaged mean field
Hamiltonian depends on thermal averaged particle density,
spin density and kinetic energy density. The thermal averaged
particle density is given by:

\beq
\rho_T(r) =\frac{1}{4\pi} \sum_{i} (2j_i+1) [ V_i^* (r) 
V_i (r) (1 - f_i ) \\
+ U_i^* (r) U_i (r) f_i ] 
\label{2}
\eeq
where $f_i = [1 + exp ( E_i/k_B T)]^{-1}$ is the 
Fermi distribution, $k_B$ is the Boltzmann constant and T is the 
temperature. For the expressions of other densities see Ref.\cite{sa04}.
 
The FT-HFB equations (\ref{1}) are solved under a spherical symmetry
assumption. The Hartree-Fock mean field is determined by using the Skyrme
force SLy4 \cite{ch98}. For the pairing interaction we choose :
$\rho_0$=0.16 fm$^{-3}$ and $\alpha$=1.  The parameter $\eta$ controls the
spatial localization of the pairing force. A volume pairing interaction
corresponds to $\eta$=0, whereas a surface type is given by $\eta$=1. The
pairing force is used with a quasiparticle energy cutoff equal to 60 MeV and
the calculations are performed in a box of radius R=18 fm.

\section{Vanishing pairing correlations in Sn and Ni isotopes}

The study is performed for the case of Sn isotopes. In what follows we
present the results of the FT-HFB calculations, first in some stable
isotopes and then for isotopes closer to the neutron-drip line.
 
Figure \ref{fig:pfield} displays the pairing field obtained in $^{124}$Sn
with surface and volume pairing interactions. In the volume pairing case,
the pairing field has similar values around the center and the surface,
with a depression in the middle of the nucleus. With a surface pairing
interaction, the pairing field exhibits a large peak in the surface of the
nucleus. To analyze how the pairing correlations are destroyed by the
temperature one usually studies the temperature dependence of the pairing gap.
Since in the present calculations the gap is state dependent, one can use as
order parameter the averaged gap. Another choice is to use the pairing
energy as order parameter. The two alternatives are compatible if the
averaged gap is defined by : 

\begin{equation}\label{eq:avpair}
\left<\Delta_n\right>_\kappa=\frac{\int {d{\bf r} \kappa_T({\bf
r})\Delta_{T,n}({\bf r})}}
{\int {d{\bf r} \kappa_T({\bf r})}}
\end{equation}
 
Figure \ref{fig:delta1} shows the thermal evolution of the mean neutron gap
in $^{124}$Sn, in the case of a surface pairing interaction. The critical
temperature above which pairing correlations vanish is T$_c$=0.7 MeV. It
should be noted that the T$_c$ $\simeq$0.5 $\Delta_{T=0}$ rule is still
qualitatively verified.

An alternative definition of the pairing gap is to use the particle density
$\rho_T$ instead of the pairing density $\kappa_T$ : 

\begin{equation}\label{eq:avpair2}
\left<\Delta_n\right>_\rho=\frac{\int {d{\bf r} \rho_T({\bf r})\Delta_{T,n}({\bf r})}}
{\int {d{\bf r} \rho_T({\bf r})}}
\end{equation}

However, in this case a too large weight is put on states located deeply
below the Fermi level where pairing effects are small. Such
a mean pairing gap is displayed in Fig. \ref{fig:delta1}. The value of the
gap at T=0 is significantly smaller with $\Delta_\rho$ than with
$\Delta_\kappa$. The volume averaged gap does not follow the BCS rule
T$_c$$\simeq$0.5 $\Delta_{T=0}$, and it is much closer to T$_c$=$\Delta_{T=0}$.
It shows that using the adequate definition $\Delta_\kappa$ of the mean
pairing gap is necessary when predicting temperature effects related to
pairing in nuclei. It should be noted that this deviation of $\Delta_\rho$
from the standard T$_c$$\simeq$0.5 $\Delta_{T=0}$ rule is only observed for
surface pairing. In the case of a volume pairing interaction, $\Delta_\rho$
and $\Delta_\kappa$ are similar and are both compatible with the
T$_c$$\simeq$0.5 $\Delta_{T=0}$ rule.

In order to compare the results obtained with volume or surface pairing
forces, one needs to give a reasonable prescription for fixing the strength
of the interaction. The best choice would be to get for both forces the same
odd-even mass difference. The alternative we have taken here was to choose,
in a given nuclei, the strength of the interactions such that to obtain the
same total binding energy. Table \ref{tabv0} summarizes the corresponding
V$_0$ values of Eq. (\ref{eqpaiint}). 

Figure \ref{fig:delta2} displays the pairing gaps in $^{104,116,124,128}$Sn
calculated with a volume and a surface pairing interaction. They generally
exhibit slightly different values at T=0 (10\% deviation). These differences
depend on the intensity of pairing correlations: V$_0^{Surf}$ in $^{124}$Sn
is somewhat smaller compared to other isotopes (Table \ref{tabv0}), whereas
V$_0^{Vol}$ doesn't change very much from $^{104}$Sn to $^{128}$Sn.
Therefore the largest differences appear in $^{104}$Sn, where in average the
pairing gaps are the largest. On the other hand, in $^{124}$Sn, in which the
averaged pairing gap was reduced compared to the other isotopes, the volume
and the surface pairing forces give practically the same energy gap at all
temperatures. The interesting fact seen in Figure \ref{fig:delta2} is that
the two gaps converge towards similar critical temperatures even in the case
when at T=0 the gaps are not the same. One can also notice that for both
surface and volume pairing forces the rule T$_c$$\simeq$ 0.5$\Delta_{T=0}$
gets approximatively fulfilled. However, for all nuclei except $^{124}$Sn
the pairing gap at T$\lesssim$ T$_c$ is significantly larger in the case of
the surface pairing interaction than in the volume one. It should be noted
that this dependence on the nature of the pairing interaction requires that
the pairing correlations at T=0 are large enough, which is the case of
$^{104,116,128}$Sn. In the recent study of Ref. \cite{sa05} it has been
shown that the pairing density depends on the angular momentum of the states
close to the Fermi level. In contrast, Fig. \ref{fig:delta2} shows that the
critical temperature is not sensitive to this feature since the nuclei under
consideration correspond to various energy positions of the Fermi level in
the N=50-82 shell.

In order to check the dependence of the critical temperature on the Skyrme
interaction, FT-HFB calculations have been performed in $^{130}$Sn with the
SGII interaction \cite{gi81} (Fig. \ref{fig:delta3}). No deviation is found
in the thermal evolution of the mean gap, compared to the above-mentioned
results obtained with the SLy4 interaction. This shows again the stability
of the critical temperature. In this case, the surface pairing interaction
gives also a larger averaged gap for T $\lesssim$ T$_c$.

The average gap results from the contributions of various single
quasiparticle states. One can have a more microscopic insight of vanishing
pairing correlations by looking at the individual pairing gap values. Fig.
\ref{fig:deltapart} displays the evolution of the pairing gaps associated
with the one quasiparticle levels of the neutron valence shell in
$^{104}$Sn, using a surface pairing interaction. The various gap values at
T=0 range from 1.3 MeV to 1.8 MeV. However, the behavior of the single
quasiparticle gaps with respect to temperature are very similar : they all
decrease so as to converge towards the same critical temperature T$_c
\simeq$ 1 MeV. Hence the critical temperature itself is independent of the
l-value of the single quasiparticle state. We have checked the validity of
this result on other nuclei such as $^{128}$Sn.

In order to investigate more deeply the role of the individual states, Fig.
\ref{fig:equpart} shows the evolution of the neutron single quasiparticle
energies with respect to the temperature in $^{104}$Sn. The relative
positions of the energy levels are given by their vicinity to the chemical
potential : the states close to the chemical potential have low single
quasiparticle energies, and are more sensitive to the pairing gap. On the
contrary, states located far from the chemical potential have larger
quasiparticle energies and are less sensitive to the pairing gap evolution.
One can see a general smooth decrease of the energies with increasing
temperature, which is due to the decrease of the pairing gap (see Fig.
\ref{fig:deltapart}). Around the critical temperature, only the lowest
single quasiparticle states such as 2d$_{5/2}$ also exhibit a strong
decrease. At higher temperatures, the single quasiparticle energies follow
the smooth increase of the single particle energies, due to the absence of
pairing correlations. This interpretation is also valid for other nuclei,
such as $^{128}$Sn.

The behavior of the vanishing of pairing correlations in exotic nuclei close
to the drip-lines is an open question. We examine in Fig. \ref{fig:delta3}
the pairing gap of the very neutron-rich nucleus $^{170}$Sn. The results are
obtained with the SLy4 interaction and the surface pairing interaction is
here adjusted to reproduce the T=0 gap value obtained with the volume type
pairing. Once again the critical temperature is predicted to be the same for
both volume and surface pairing interactions : T$_c$=0.8 MeV. This result
stresses again the strong stability of the critical temperature with respect
to pairing localization properties, also in very neutron-rich nuclei. There
is, however, a difference in the thermal evolution of the two mean gaps in
$^{170}$Sn, as noticed in previous Sn isotopes. For instance, at T=0.7 MeV
the gap is two times higher for a surface pairing interaction than for a
volume one. Hence pairing correlations at temperatures below T$_c$ in
neutron-rich as well as in stable nuclei may provide information about the
surface or volume type of the pairing interaction. 

Calculations have also been performed in Ni isotopes, and the results are
illustrated here for the case of the $^{84}$Ni nucleus. The T$_c$$\simeq$
0.5$\Delta_{T=0}$ rule is still verified, both with surface and volume type
interactions. The surface and the volume pairing interactions lead to
similar critical temperatures, with 100 keV variation between the two cases.
The situation is analogous in $^{104}$Sn (Fig. \ref{fig:delta2}) where the
surface and volume interactions give the largest variation in T$_c$, namely
100 keV. This upper limit shows that the absolute value of the critical
temperature itself is rather independent of the nature of the pairing
interaction. However, if measurements can reach such a resolution it would
be of high interest to measure pairing properties at temperatures located
just below T$_c$ : for instance, in $^{104}$Sn at T=0.8 MeV, the pairing gap
is close to zero in the volume case whereas the gap remains $\Delta$=1 MeV
in the surface case. This is also the case for $^{84}$Ni. 

Fig. \ref{fig:cv} displays the specific heat for $^{84}$Ni,
obtained in the FT-HFB calculations.
%, i.e. :
%\begin{equation}
%C_V=\frac{\partial E_{tot}}{\partial T}
%\end{equation}
Due to the finite step in temperature used in the HFB calculations, the
specific heat displays a kink at the critical temperature instead of the
usual singularity. Actually, due to the finite size of the nucleus, the
specific heat should have a smooth $s$ shape behavior around the critical
temperature \cite{sc01}. In order to get this behavior in the calculations,
one needs to go beyond the HFB approach, e.g., projecting out the number of
particles and taking into account the thermal fluctuations \cite{li01}.

\section{Conclusion}

In conclusion, the critical temperature for vanishing of pairing
correlations in hot nuclei appears to be rather insensitive to the surface
or volume localization of the pairing force used in FT-HFB calculations.
For all the neutron-rich tin isotopes studied, we have found that the
critical temperature is given approximatively by T$_c$$\simeq$0.5
$\Delta_{T=0}$ for both type of pairing forces and for stable and unstable
nuclei. Hence, the critical temperature could be deduced from the gap value
at zero temperature. On the other hand, the pairing gap is strongly
sensitive to the nature of the pairing interaction for temperatures just
below the vanishing of pairing correlations, for the large majority of
nuclei studied. This result should open an experimental investigation for
the pairing interaction in nuclei.

\newpage
\begin{table}[h]
\begin{tabular}{|c||c|c|c|c|c|c|c|}
%\\ \hline
%\\ \cline{1-5}
  & \textbf{$^{104}$Sn}   & \textbf{$^{116}$Sn} & \textbf{$^{124}$Sn}  & \textbf{$^{128}$Sn} &
  \textbf{$^{130}$Sn} & \textbf{$^{170}$Sn} &\textbf{$^{84}$Ni} 
\\ \hline
 V$_0$$^{Vol}$ (MeV.fm$^3$) & -220 & -200 & -197 & -220 & -220 & -220 & -240  
\\ \hline
 V$_0$$^{Surf}$ (MeV.fm$^3$)  & -580 & -520 & -490 & -570 & -575 & -510 & -502  
\\ \hline
\end{tabular}
\caption{\label{tabv0}}
V$_0$ values of Eq. (\ref{eqpaiint}) corresponding to volume
and surface pairing interactions, for the different nuclei considered
\end{table}

\begin{figure}[h]
\vspace{0.0cm}
\caption {Pairing field at T=0 in $^{124}$Sn, calculated with the HFB approach and
the SLy4 interaction. Solid and dashed lines correspond to volume and
surface interactions, respectively.
}
\label{fig:pfield}
\end{figure}

\begin{figure}[h]
\vspace{0.0cm}
\caption {Mean value of the neutron pairing gap in $^{124}$Sn, calculated
with Eq. (\ref{eq:avpair}) (solid line) and Eq. (\ref{eq:avpair2}) (dashed line).} 
\label{fig:delta1}
\end{figure}

\begin{figure}[h]
\vspace{0.0cm}
\caption {Mean value of the neutron pairing gap in $^{104,116,124,128}$Sn, calculated
with a volume pairing interaction (solid line) and a surface one (dashed line).}
\label{fig:delta2}
\end{figure}

\begin{figure}[h]
\vspace{0.0cm}
\caption {Mean value of the neutron pairing gap in $^{130}$Sn (with SGII
interaction) and $^{170}$Sn (with SLy4 interaction), calculated
with a volume pairing interaction (solid line) and a surface one (dashed line)}
\label{fig:delta3}
\end{figure}

\begin{figure}[h]
\vspace{0.0cm} \caption {Neutron pairing gaps in $^{104}$Sn corresponding to
the quasiparticle states of the N=50-82 valence shell.}
\label{fig:deltapart}
\end{figure}

\begin{figure}[h]
\vspace{0.0cm} \caption {Neutron single quasiparticle energies  in $^{104}$Sn corresponding to
the states of the N=50-82 valence shell.}
\label{fig:equpart}
\end{figure}

\begin{figure}[h]
\vspace{0.0cm}
\caption {Specific heat in $^{84}$Ni calculated with a surface pairing
interaction.}
\label{fig:cv}
\end{figure}

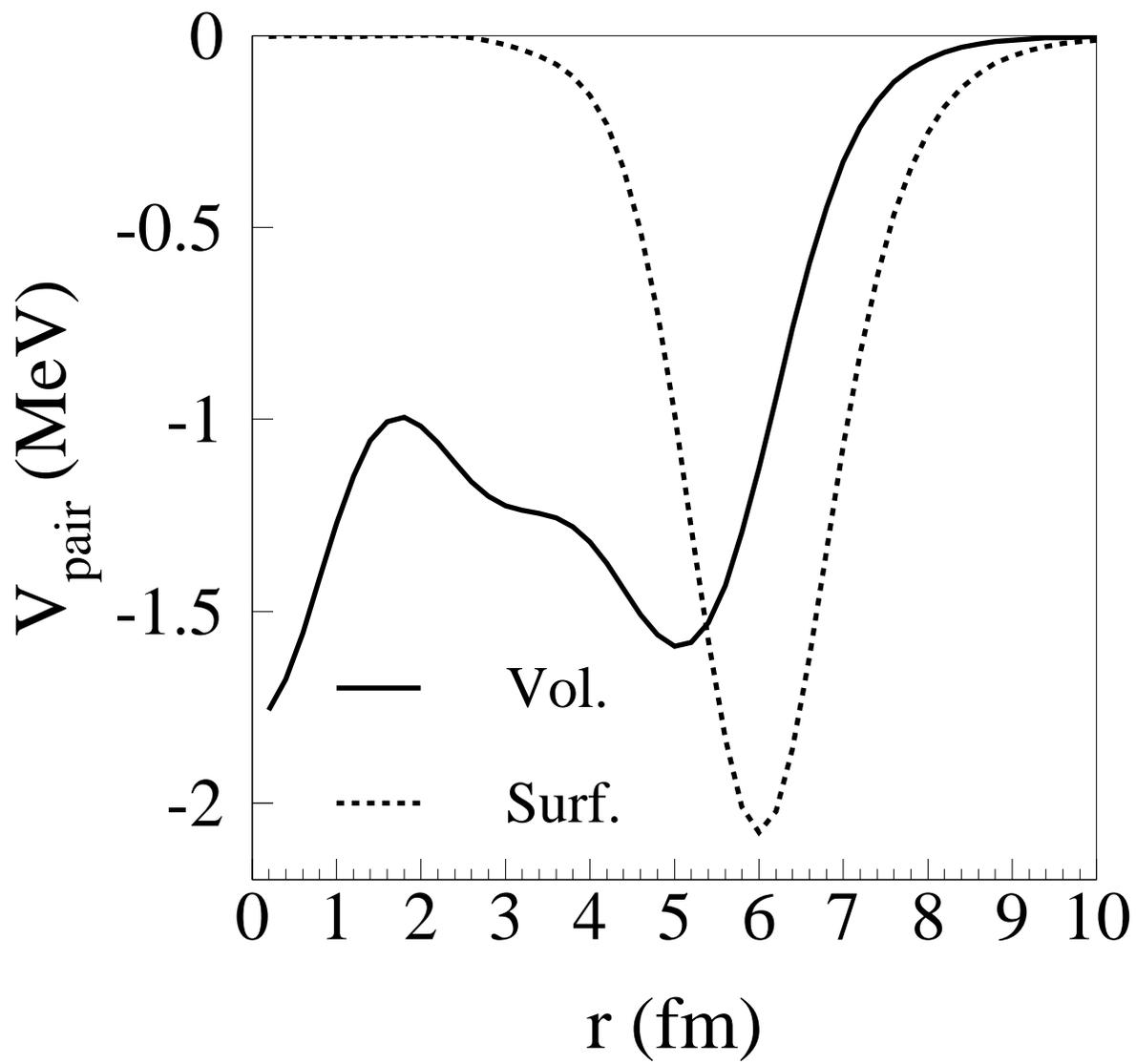
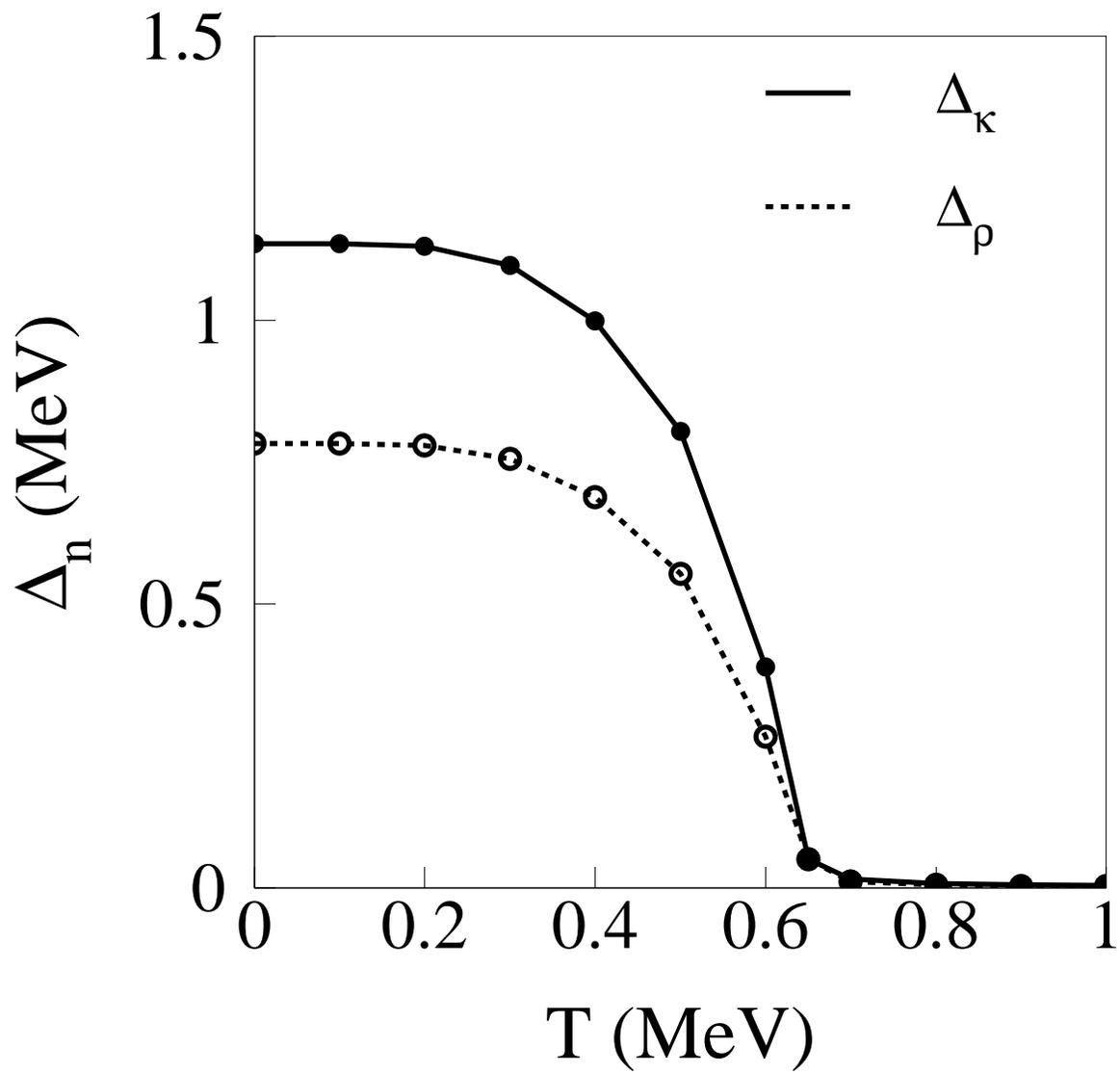
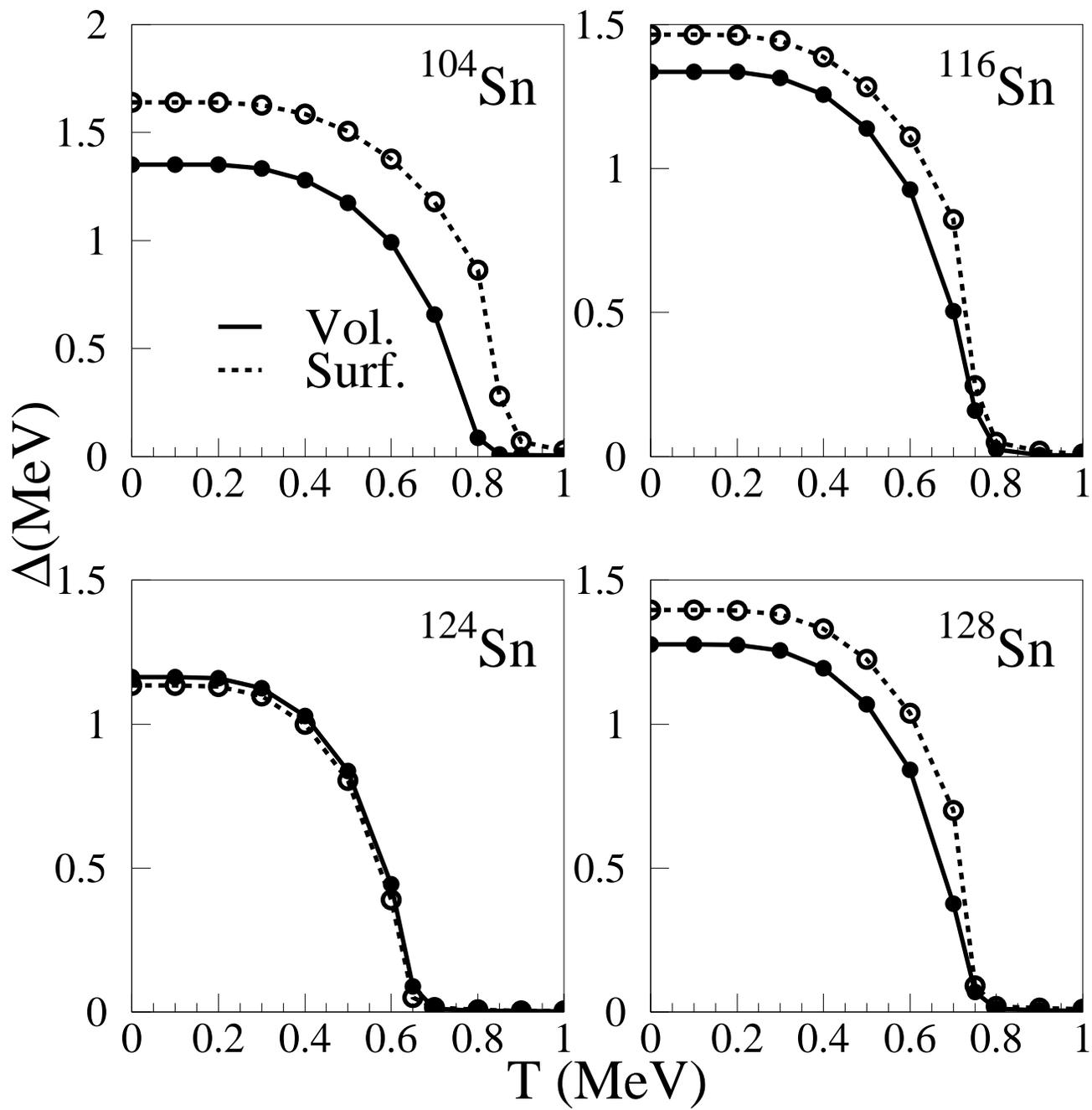
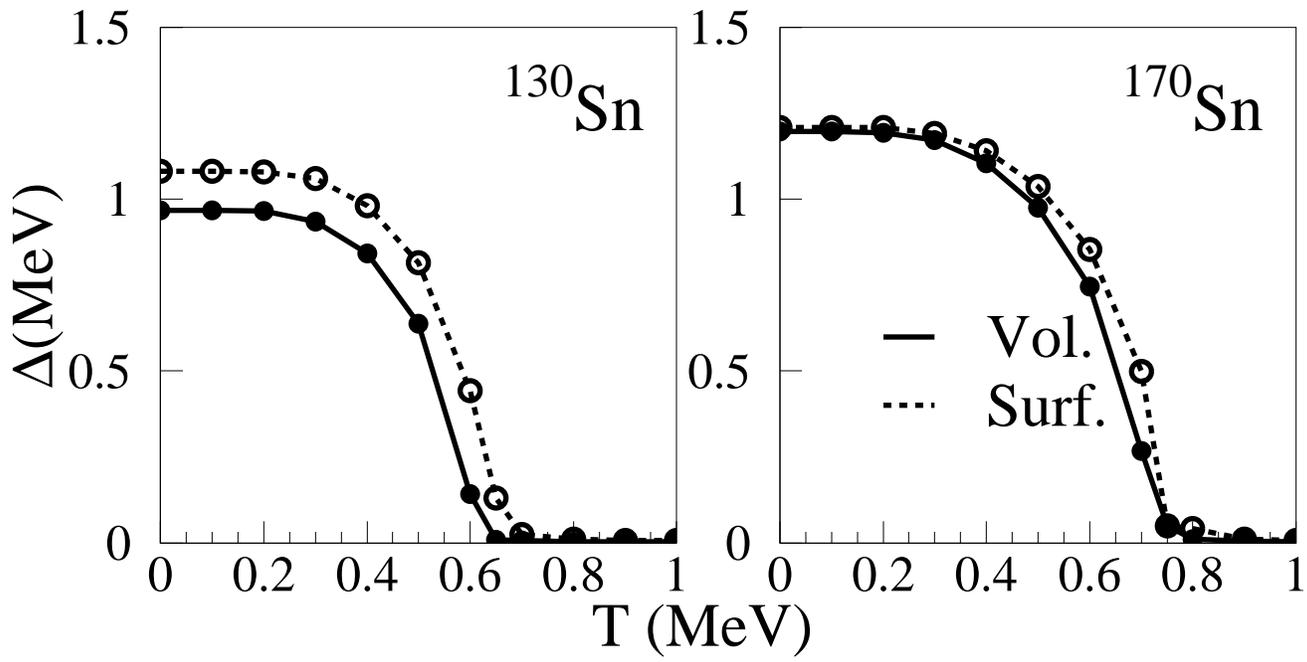
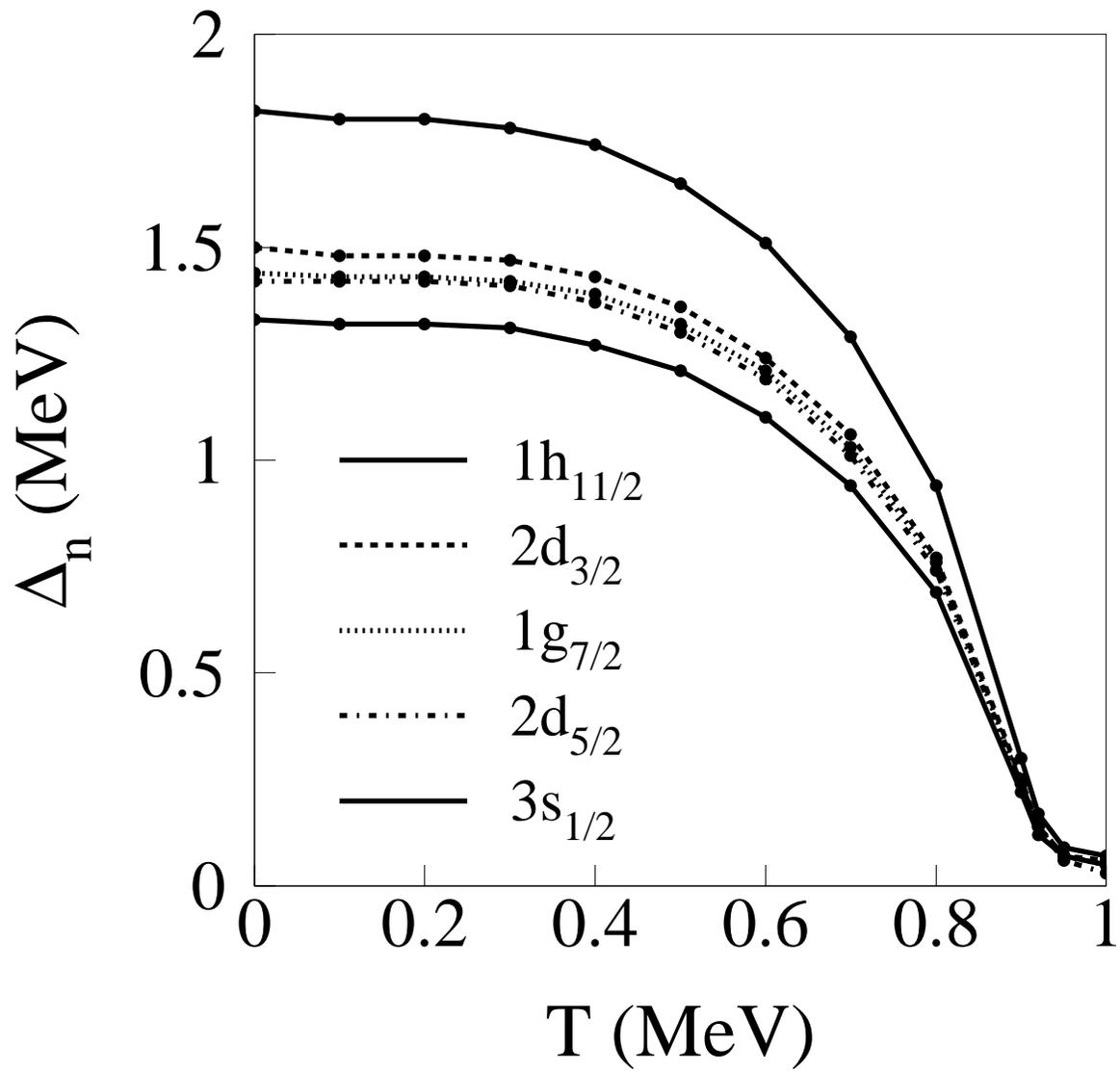
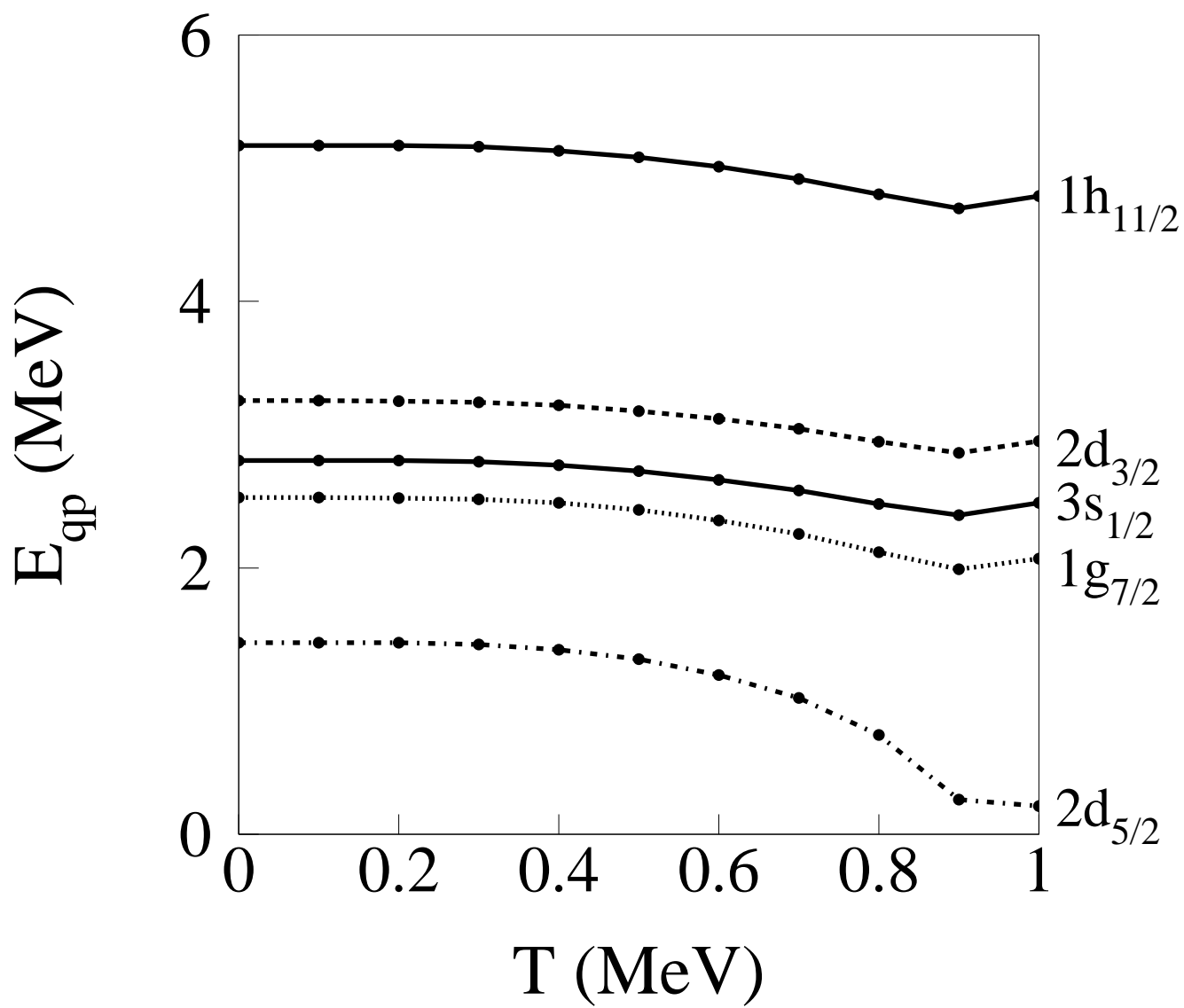
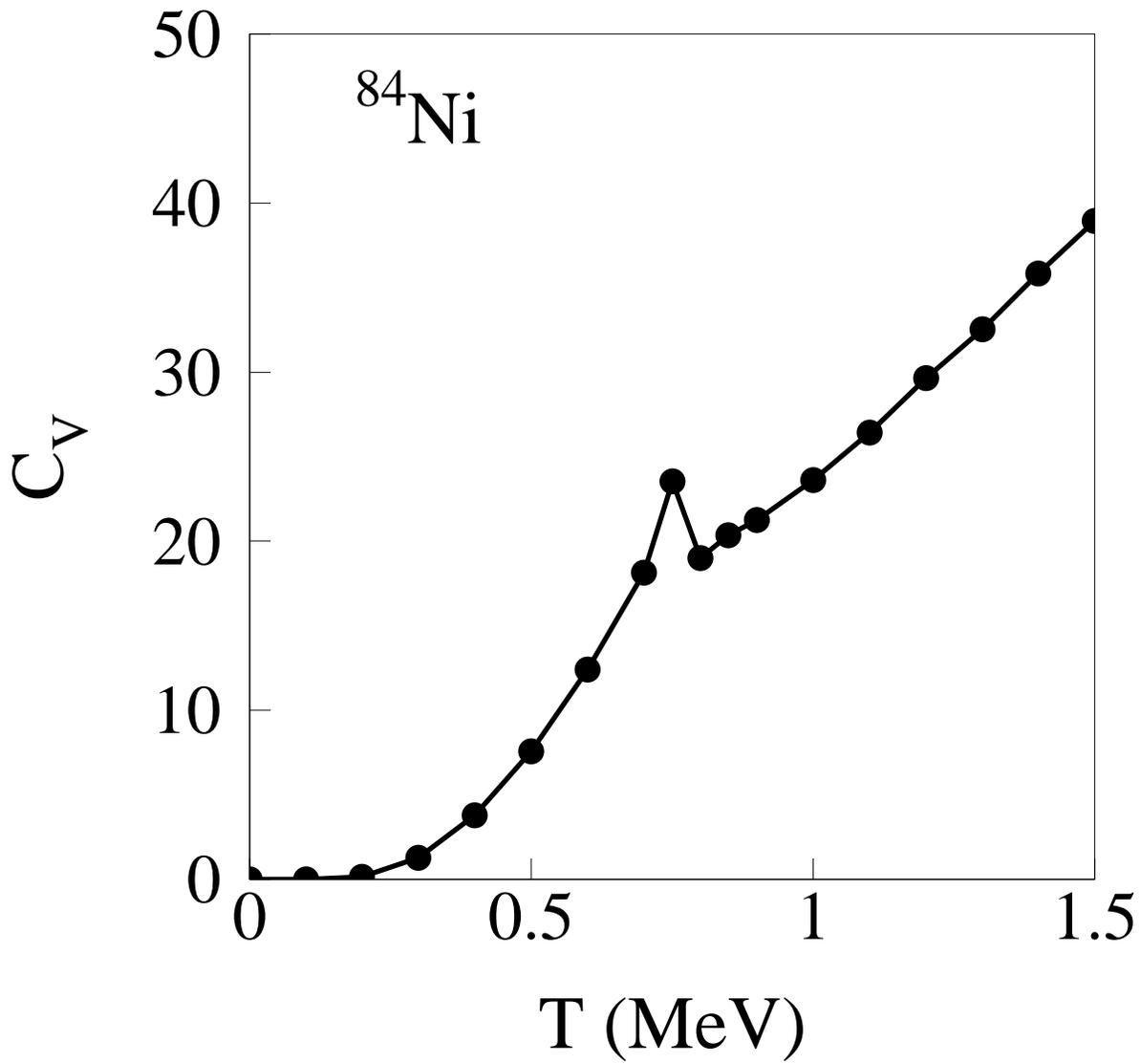
\end{document}